\documentclass[newstyle,twocolumn,proceedings]{rmaa}
\title{Effects of CSPN Models on PNe Shell Modeling}
\suppressfulladdresses
\author{B. Armsdorfer\altaffilmark{1}, S. Kimeswenger\altaffilmark{1},
and T. Rauch\altaffilmark{2} \\
birgit.armsdorfer@uibk.ac.at, stefan.kimeswenger@uibk.ac.at, rauch@astro.uni-tuebingen.de
 \altaffiltext{1}{Institut f{\"u}r Astrophysik, Universit{\"a}t Innsbruck,
Austria.}
 \altaffiltext{2}{Institut f{\"u}r Astronomie und Astrophysik, Universit{\"a}t T\"ubingen, Germany.}}
\shortauthor{Armsdorfer et al.}
\shorttitle{Effects of CSPN Models on PNe Shell Modeling}
\listofauthors{B.~Armsdorfer, S.~Kimeswenger, \& T.~Rauch}
\indexauthor{Armsdorfer, B.}
\indexauthor{Kimeswenger, S.}
\indexauthor{Rauch, T.}
\begin{document}
\maketitle
 
\noindent\textbf{Modeling planetary nebulae (PNe) shells using different
ionizing spectra for a hot central star (CSPN) of evolved PNe, we found that especially 
a blackbody model leads to wrong nebular diagnostics and
abundances.}

\begin{figure}[!t]
 \includegraphics[width=\columnwidth]{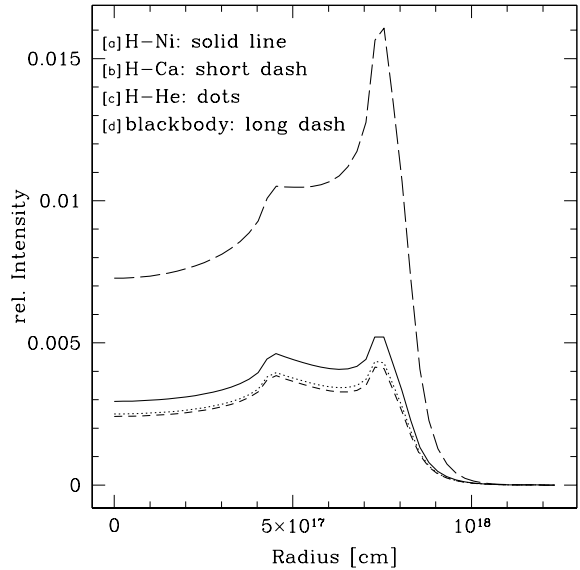}
  \caption{Results of our PNe shell modeling from the line structure
output of Cloudy for \ion{He}{2}$_{4686}$, the lines are normalized to the integrated H$\beta$ 
flux.
The blackbody model {\tt [d]} overestimates the flux, the changes
from model {\tt [a]} to  model {\tt [c]} are also visible.}
    \label{fighe}
\end{figure}

The aim of our investigations is to combine modeling of stellar
atmospheres and of planetary nebulae shells to obtain self consistent
descriptions of the nebulae. 
For this purpose, spatially resolved roundish nebulae with well modeled central stars were selected
(Rauch et al.\ 1999). As a test sample, the nebula NGC 2438 was
selected, and the photo-ionization code Cloudy  (Ferland 1996) 
was used for the modeling.
To tune nebular parameters like density profile and filling factor,
we used one of the new NLTE stellar model atmosphere fluxes 
with H-Ni element coverage by Rauch (2001) with $T_{\rm eff}$=110 kK and $\log\rm g$=7 (cgs).
The other geometrical parameters were derived from
narrow-band images and long-slit spectra. This model was then used to calculate four different 
sets of artificial narrow-band surface brightness distributions in
various species (\ion{H}{1}, \ion{He}{1}, \ion{He}{2}, \ion{O}{3},
\ion{N}{2}, \ion{S}{2}).
The sources of illumination were {\tt [a]} the new NLTE model
atmosphere by Rauch as mentioned above, {\tt [b]} a H-Ca NLTE model (Rauch 1997), 
{\tt [c]} a NLTE atmosphere containing only H-He, and {\tt [d]} a
blackbody. 
The stellar fluxes were calibrated to have the same visual magnitude,
$T_{\rm eff}$ and $\log\rm g$. While H$\alpha$ and H$\beta$ are nearly unaffected by the different
illumination sources, other lines change both in 
strength and spatial profile. For the low ionized species, the
blackbody overestimates the lines by 30\%
([\ion{S}{2}]) to 45\%
([\ion{N}{2}])
-- 
therefore, also the ratio H$\alpha$/[\ion{S}{2}] vs.
H$\alpha$/[\ion{N}{2}] used for PNe identification diagrams
changes up to 15\%. 
For the [\ion{O}{3}] lines,
the blackbody underestimates the real flux by 30\%.
 \ion{He}{1}$_{4471}$ is underestimated by
20\%, while \ion{He}{2}$_{4686}$, which also shows the strongest spatial
change,  is overestimated by 270\%  
with the blackbody model (see Fig.~\ref{fighe}). 
Models {\tt [b]} and {\tt [c]} underestimate the
flux by about 20\%.
All diagnostic diagrams are therefore affected. This leads typically to
an overestimation of the stellar temperature and nitrogen
abundance.
A blackbody model for the stellar
atmosphere thus leads to wrong nebular parameters and should not be used
anymore.\\
This work will be discussed more detailed in a forthcoming paper.

\acknowledgments
This  project  was  supported by the BMBWK, Sektion VIII/A/5, and by a
DLR grant 50 OR97055.
Support from the organizers to participate in the conference is
gratefully acknowledged.

\end{document}